\documentclass[sigconf]{acmart}
\usepackage{kotex}
\usepackage{adjustbox}
\usepackage{subfigure}
\usepackage{framed,multirow}

\AtBeginDocument{%
  \providecommand\BibTeX{{%
    \normalfont B\kern-0.5em{\scshape i\kern-0.25em b}\kern-0.8em\TeX}}}

\newcommand\HH{
  \global\let\savedtextbullet\textbullet
  \gdef\textbullet{%
    \par\noindent\savedtextbullet\global\let\textbullet\savedtextbullet
  }%
}

\copyrightyear{2021}
\acmYear{2021}
\setcopyright{acmcopyright}\acmConference[SIGIR '21]{Proceedings of the 44th International ACM SIGIR Conference on Research and Development in Information Retrieval}{July 11--15, 2021}{Virtual Event, Canada}
\acmBooktitle{Proceedings of the 44th International ACM SIGIR Conference on Research and Development in Information Retrieval (SIGIR '21), July 11--15, 2021, Virtual Event, Canada}
\acmPrice{15.00}
\acmDOI{10.1145/3404835.3463076}
\acmISBN{978-1-4503-8037-9/21/07}

\begin{document}
\fancyhead{}

\title{Improving Bi-encoder Document Ranking Models with Two Rankers and Multi-teacher Distillation}

\author{Jaekeol Choi}
\affiliation{%
	\institution{Seoul National University \\ \& Naver Corp.}
}
\email{jaekeol.choi@snu.ac.kr}


\author{Euna Jung, Jangwon Suh}

\affiliation{%
	\institution{GSCST \\ 
	Seoul National University}
	\streetaddress{}
}
\email{xlpczv@snu.ac.kr, rxwe5607@snu.ac.kr}

\author{Wonjong Rhee}
\affiliation{%
	\institution{GSCST, GSAI, AIIS \\
	Seoul National University}
}
\email{wrhee@snu.ac.kr}



\begin{abstract}
BERT-based Neural Ranking Models (NRMs) can be classified according to how the query and document are encoded through BERT's self-attention layers - \textit{bi-encoder} versus \textit{cross-encoder}.
Bi-encoder models are highly efficient because all the documents can be pre-processed before the query time, but their performance is inferior compared to cross-encoder models.
Both models utilize a ranker that receives BERT representations as the input and generates a relevance score as the output.
In this work, we propose a method where multi-teacher distillation is applied to a cross-encoder NRM and a bi-encoder NRM to produce a bi-encoder NRM with two rankers.
The resulting student bi-encoder achieves an improved performance by simultaneously learning from a cross-encoder teacher and a bi-encoder teacher and also by combining relevance scores from the two rankers. 
We call this method TRMD (Two Rankers and Multi-teacher Distillation).
In the experiments, TwinBERT and ColBERT are considered as baseline bi-encoders. 
When monoBERT is used as the cross-encoder teacher, together with either TwinBERT or ColBERT as the bi-encoder teacher, TRMD produces a student bi-encoder that performs better than the corresponding baseline bi-encoder. For P@20, the maximum improvement was 11.4\%, and the average improvement was 6.8\%. 
As an additional experiment, we considered producing cross-encoder students with TRMD, and found that it could also improve the cross-encoders.\footnote{The code is available at https://github.com/maygodwithu/TRMD.git}
\end{abstract}




\maketitle


\section{Introduction}
During the past few years, a variety of neural ranking models (NRMs) based on BERT have been suggested in the information retrieval (IR) community. 
Starting with monoBERT \cite{nogueira2019passage}, a number of document ranking models \cite{macavaney2019cedr,lu2020twinbert,khattab2020colbert} have adopted BERT for NRMs.
These BERT-based models tend to be robust against the vocabulary mismatch problem because they learn semantic representations of query–document pairs in a latent space \cite{mitra2018introduction}. 

\begin{table}[h]
	\caption{BERT-based document ranking algorithms.}
	\label{tab:alg_class}
	\adjustbox{max width=\linewidth}{%
	\begin{tabular}{ccl}
		\toprule
		Algorithm & BERT encoder & BERT representation\\
		\midrule
		monoBERT \cite{nogueira2019passage} & \textit{cross-encoder} & \textit{CLS} \\
		CEDR-KNRM \cite{macavaney2019cedr} & \textit{cross-encoder} & \textit{CLS},\textit{query/document} \\
		TwinBERT \cite{lu2020twinbert} & \textit{bi-encoder} & \textit{query/document} \\
		ColBERT \cite{khattab2020colbert} & \textit{bi-encoder} & \textit{CLS},\textit{query/document} \\
		\bottomrule
	\end{tabular}}
\end{table}

\begin{figure*}[t]
	\centering
	\includegraphics[width=0.95\linewidth]{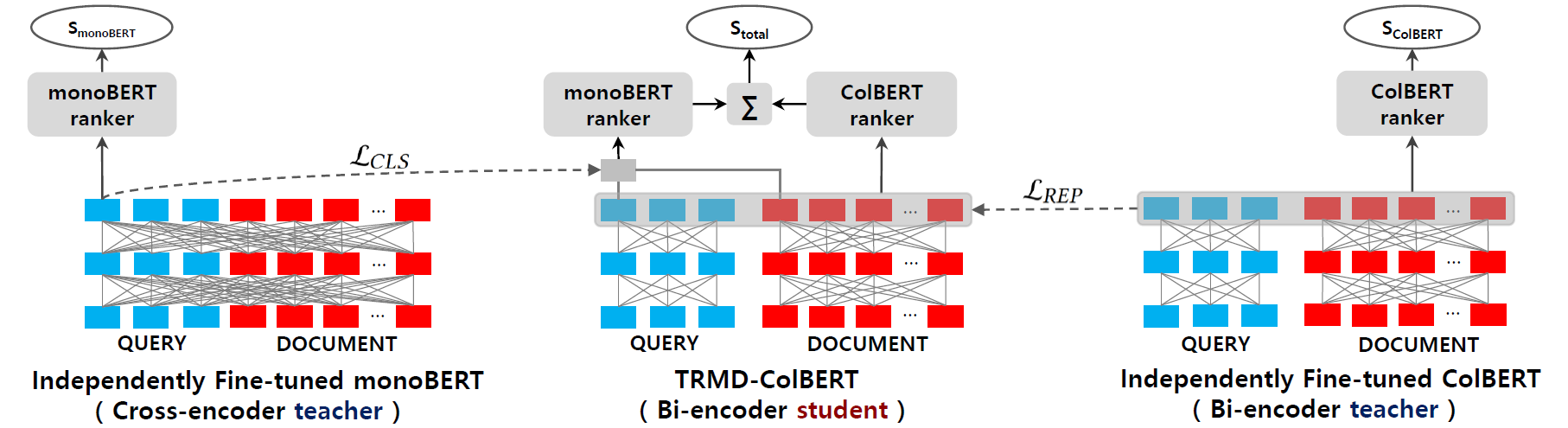}
	\caption{
	Architecture of TRMD-ColBERT. (\textit{left}) Fine-tuned monoBERT is used as a cross-encoder teacher. monoBERT is used to distill the knowledge from its CLS representation to the student.
	(\textit{middle}) TRMD-ColBERT estimates the score by combining scores from two rankers, each of which uses the same representations as those used by each teacher ranker. TRMD-ColBERT is trained by distilling the knowledge from the representation of each teacher to the student.
	(\textit{right}) Fined-tuned ColBERT is used as a bi-encoder teacher. ColBERT is used to distill knowledge from its representations to the student. }
	\label{fig:duet_distil}
	\Description{}
\end{figure*}

We categorize BERT-based NRMs according to BERT encoder type and BERT representation usage.
First, following Humeau et al.\cite{humeau2019poly}, we classify BERT-based NRMs into `\textit{cross-encoder}' and `\textit{bi-encoder}' models. \textit{Cross-encoder} models perform a full self-attention over the entire query-document pair, whereas \textit{bi-encoder} models perform two independent self-attentions over the query and the document.
The BERT layers of \textit{cross-encoder} NRMs can model the interaction between a query and a document, and the resulting BERT representations contain contextualized embedding. On the other hand, the BERT self-attention layers of \textit{bi-encoder} NRMs cannot model the interaction. Therefore, a document is mapped to a fixed BERT representation regardless of the choice of a query. This makes it possible for bi-encoder models to pre-compute document representations offline, significantly reducing the computational load per query at the time of inference. In general, \textit{cross-encoder} models tend to perform better than \textit{bi-encoder} models but at the expense of a higher computational cost.
Second, we classify BERT-based NRMs models into `\textit{CLS}', `\textit{query/document}', and `\textit{CLS,query/document}', 
according to which BERT output representations are used for the relevance score calculation. The known BERT-based NRMs use one of the three representation choices.
For example, monoBERT uses only CLS representation for calculating the relevance score, whereas ColBERT uses both CLS and query/document representations.
Table \ref{tab:alg_class} summarizes four of the well-known algorithms with their categorization according to BERT encoder type and BERT representation usage.

To improve the performance of bi-encoder models that are computationally efficient, we propose a method of employing a two-ranker structure and multi-teacher distillation \cite{hinton2015distilling, fukuda2017efficient}. The method is named as \textbf{TRMD (Two Rankers and Multi-teacher Distillation)}. 
Here, the word `ranker' refers to the post-BERT part of the NRM model where BERT representations are used as the ranker's input, and the relevance score is calculated as the ranker's output. Examples can be found in Figure \ref{fig:duet_distil}. 
With TRMD, the student model utilizes two rankers. One ranker is assigned to use the representation that is distilled from a high-performance cross-encoder as the input. The other ranker is assigned to use the representation that is distilled from an efficiently-structured bi-encoder as the input. Obviously, the encoders become the two teachers.


In our experiments, monoBERT \cite{nogueira2019passage} is chosen as the cross-encoder teacher, while either TwinBERT \cite{lu2020twinbert} or ColBERT \cite{khattab2020colbert} is chosen as the bi-encoder teacher. As shown in the third column of Table \ref{tab:alg_class}, different models use different parts of BERT representations for estimating the relevance score. TRMD forces the student's BERT representations to become similar to the respective teacher's BERT representations.
An example is shown in Figure \ref{fig:duet_distil} where monoBERT and ColBERT are used as the two teachers. In this case, the bi-encoder student learns both of monoBERT's CLS representation and ColBERT's CLS and query/document representations. 
As will be shown in the experiment section, the results on Robust04 and Clueweb09 datasets show that TRMD can significantly improve the performance of the existing bi-encoder models: TwinBERT and ColBERT.

\section{Methodology}
Figure \ref{fig:duet_distil} shows the architecture of TRMD-ColBERT in the middle, where TRMD-ColBERT is the name of a bi-encoder student that is produced by improving ColBERT using TRMD. When TwinBERT is improved using TRMD, the resulting bi-encoder is called TRMD-TwinBERT. While there is no restriction on what model to use as the cross-encoder teacher, we only consider monoBERT in this work, and therefore the naming does not reflect the choice of the cross-encoder.
In this section, we illustrate how TRMD process works when monoBERT and ColBERT are employed as the two teachers.

\subsection{Architecture}
As shown in Figure \ref{fig:duet_distil}, TRMD-ColBERT combines two rankers on top of the BERT encoder such that it can learn from two different teachers.
With this structural modification, the bi-encoder student (ColBERT) can directly integrate the knowledge of both teachers (monoBERT and ColBERT).
Each ranker in the resulting TRMD-ColBERT uses the same BERT representation as what the ranker of its respective teacher uses.

While ColBERT has two CLS representations (for query and document), a ranker in monoBERT uses only one CLS representation that reflects query-document interaction. Hence, adding monoBERT ranker to ColBERT is not straightforward. To address this issue, we combine two CLS representation vectors of ColBERT into one vector. Following TwinBERT \cite{lu2020twinbert}, we apply residual function \cite{he2016deep} to combine the two vectors, where the formal definition of the residual function is as follows.
\begin{equation}
\label{eq:residual_network} 
\textit{CLS} = F(x,W,b)+x
\end{equation}
Here, $x$ is the max of the query's and document's CLS vectors and $F$ is a linear mapping function from $x$ to the residual with parameters $W$ and $b$.
A student's ranker corresponding to monoBERT uses the resulting CLS representation to estimate the score $S_{monoBERT}$.
TRMD-ColBERT also utilizes a ColBERT ranker for relevance estimation, producing $S_{ColBERT}$. A ranker corresponding to ColBERT uses the same method to calculate the relevance score as in the original ColBERT teacher. The total score of TRMD-ColBERT, $S_{total}$, is obtained by adding the two scores from the two rankers.
\begin{equation}
\label{eq:score_sum}
S_{total} = S_{monoBERT} + S_{ColBERT}
\end{equation}

TRMD-TwinBERT can be constructed in the same manner as in constructing TRMD-ColBERT. 
One notable point is that TwinBERT ranker can use either \textit{CLS} representation or \textit{query/document} representation.
In our experiment, we chose TwinBERT using \textit{query/document} representation to make TwinBERT ranker use a different representation from monoBERT's choice of representation.

\subsection{Learning through Multi-teacher Distillation}
Knowledge distillation \cite{hinton2015distilling} has been used to train a small but high-performing model by forcing a small student model to learn from a large teacher model or from multiple teachers \cite{you2017learning, fukuda2017efficient}. TRMD trains a student model by distilling knowledge from two teachers with different BERT encoder types, a cross-encoder and a bi-encoder.
As shown in Figure \ref{fig:duet_distil}, monoBERT and ColBERT become the teachers, a single bi-encoder ColBERT becomes the student model, and each teacher distills the knowledge contained in its BERT output representation to the student's representation.
Before applying knowledge distillation, we independently fine-tune the pre-trained teacher models such that their individual performance is optimized.

During the training of the student model with knowledge distillation, we introduce three loss terms.
First, the hard prediction loss $\mathcal{L}_{hard-pred}$ is defined as the hinge loss between the student's prediction score and the true relevance score of the document, and it is shown in Equation (\ref{eq:triplet}).
Second, \textit{CLS} loss $\mathcal{L}_{CLS}$ encourages the student model to learn the \textit{CLS} representation from the monoBERT teacher using MSE as in Equation (\ref{eq:cls}). Third, representation loss $\mathcal{L}_{REP}$ forces the student's representations to resemble the representations of ColBERT using MSE between the two representations as in Equation (\ref{eq:rep}).
\begin{equation}
\label{eq:triplet}
\mathcal{L}_{hard-pred} = \text{Hingeloss}(\text{softmax}(z^{S}))
\end{equation}
\begin{equation}
\label{eq:cls}
\mathcal{L}_{CLS} = MSE({CLS}^{T},{CLS}^{S})
\end{equation}
\begin{equation}
\label{eq:rep}
\mathcal{L}_{REP} = MSE({REP}^{T},{REP}^{S})
\end{equation}

In Equations (\ref{eq:triplet}), (\ref{eq:cls}), and (\ref{eq:rep}), $z$ is a vector whose elements are the predicted scores of a positive document and a negative document, $CLS$ stands for a \textit{CLS} representation vector of BERT, $REP$ is the whole representation generated by BERT including CLS, query and document representation vectors, and $(\cdot)^{T}$ and $(\cdot)^{S}$ indicate the teacher model and the student model, respectively. 
We define the total loss for the distillation process as the sum of these three losses as in Equation (\ref{eq:final_loss}). 
%
\begin{equation}
\label{eq:final_loss}
\mathcal{L}_{total}=\mathcal{L}_{hard-pred}+\mathcal{L}_{CLS}+\mathcal{L}_{REP}
\end{equation}

\section{Experimental Result}

 \begin{table*}[t]
	\caption{
	Re-ranking performance on Robust04 and Clueweb09b. TR stands for Two Rankers, and TRMD stands for Two Rankers and Multi-teacher Distillation.}
	\label{tab:result0}
	\adjustbox{max width=\textwidth}{%
	\begin{tabular}{rc|lcc|ccc}
		\toprule
		\multirow{2}{*}{Model} & \multirow{2}{*}{BERT encoder} &
		\multicolumn{3}{c|}{\textbf{Robust04}} & \multicolumn{3}{c}{\textbf{Clueweb09b}} \\
		& & P@20 & NDCG@5 & NDCG@20 & P@20 & NDCG@5 & NDCG@20 \\
		\midrule
		monoBERT & \textit{cross} & 0.39320 & 0.53996 & 0.47874 & 0.30964 & 0.32912 & 0.30176 \\
		\midrule
		TwinBERT & \textit{bi} & 0.30172 & 0.38018 & 0.34766 & 0.21092 & 0.18074 & 0.19034 \\
		TR-TwinBERT & \textit{bi} & 0.30208 & 0.37270 & 0.34434 &  0.21004 & 0.17686 & 0.18522 \\
		TRMD-TwinBERT & \textit{bi} & \bf{0.33600} & \bf{0.40950} & \bf{0.37930} & \bf{0.22056} & \bf{0.18668} & \bf{0.19106} \\
		\midrule
		ColBERT & \textit{bi} & 0.32542 & 0.40312 & 0.37540 & 0.25074 & 0.22730 & 0.23648 \\
		TR-ColBERT & \textit{bi} & 0.31958 & 0.36180 & 0.35254 & 0.24818 & 0.21682 & 0.22674 \\
		TRMD-ColBERT & \textit{bi} & \bf{0.34500} & \bf{0.42692} & \bf{0.39452} & \bf{0.26418} & \bf{0.25672} & \bf{0.25312} \\
	   \midrule
	   \midrule
	   TRMD-TwinBERT & \textit{cross} & 0.40226 & 0.54454 & 0.47376 & \bf{0.32352} & \bf{0.33736} & \bf{0.31362} \\
	   TRMD-ColBERT & \textit{cross} & \bf{0.40384} & \bf{0.54962} & \bf{0.47611} & 0.30992 & 0.32796 & 0.30132 \\
       \bottomrule
	\end{tabular}}
\end{table*}

The goals of the experiment are as the following.
First, we would like to confirm that a performance gap exists between cross-encoder NRMs and bi-encoder NRMs.
Second, we demonstrate that the teachers in TRMD can effectively distill their knowledge to the student by examining the loss values during training.
Third, we show that applying TRMD significantly improves the performance of TwinBERT and ColBERT.
Finally, with an ablation study, we investigate if the improvements mainly come from the simple architectural change of using two rankers or from the knowledge distillation.

\subsection{Experimental Setup}
\subsubsection{Datasets and Metrics}
We conduct our experiments on Robust04 and WebTrack 2009 datasets as in \cite{macavaney2019cedr}. Following \cite{huston2014parameters}, we divide each data into five folds and use three folds for training, one for validation, and the remaining one for test. We use document collections from TREC discs 4 and 5\footnote{520k documents, https://trec.nist.gov/data-disks.html} of Robust04 and from ClueWeb09b\footnote{50M web pages, https://lemurproject.org/clueweb09/} of WebTrack 2009. For evaluation metrics, we used P@20, nDCG@5, and nDCG@20.

\subsubsection{Implementation}
Our models are implemented with Python 3 and PyTorch 1. We use the popular transformers\footnote{https://github.com/huggingface/transformers} library for the pre-trained BERT model. We used 1$\sim$2 GPUs each of which is RTX2080ti with 11G memory in parallel.

\subsubsection{Training and optimization}
To train BERT-based models, we generate data in a triplet form. A sample of a triplet consists of a query, a positive document, and a negative document. A positive(negative) document is arbitrarily selected from top-k positive(negative) documents filtered by BM25. The batch size for training is 32. 
We use pairwise hinge loss \cite{dehghani2017neural} that is known to work well with triplet data.
All parameters other than BERT parameters are trained using the Adam \cite{kingma2015adam} optimizer with a learning rate of 0.001 for 50 epochs, while the BERT parameters are trained at a learning rate of 1e-6 following the prior work \cite{hui2018co}. Documents are truncated to have up to 800 tokens. For validation, P@20 was used.

In the training process, we first train monoBERT, TwinBERT, and ColBERT independently, and then we train our student model while the weights of the teacher BERT models are frozen. As our GPU can handle only one BERT model at a time owing to the memory capacity, we use multi-GPUs in parallel using the \textit{model parallel} provided by PyTorch. Throughout the training, we feed the same training and validation data to both teacher models in each step.

\subsubsection{Baseline Models}
For TRMD, we use three BERT-based models - monoBERT, TwinBERT, and ColBERT - that are also used as the baseline models. Besides TRMD, we additionally experiment models with two rankers, but without any distillation from the teachers, to investigate the effect of the suggested distillation method. These models are used for ablation study, and they are called \textbf{TR (Two Rankers)} in contrast to TRMD.

\begin{figure}[t]
    \centering
    \subfigure[TRMD-TwinBERT]{\includegraphics[width=0.48\linewidth]{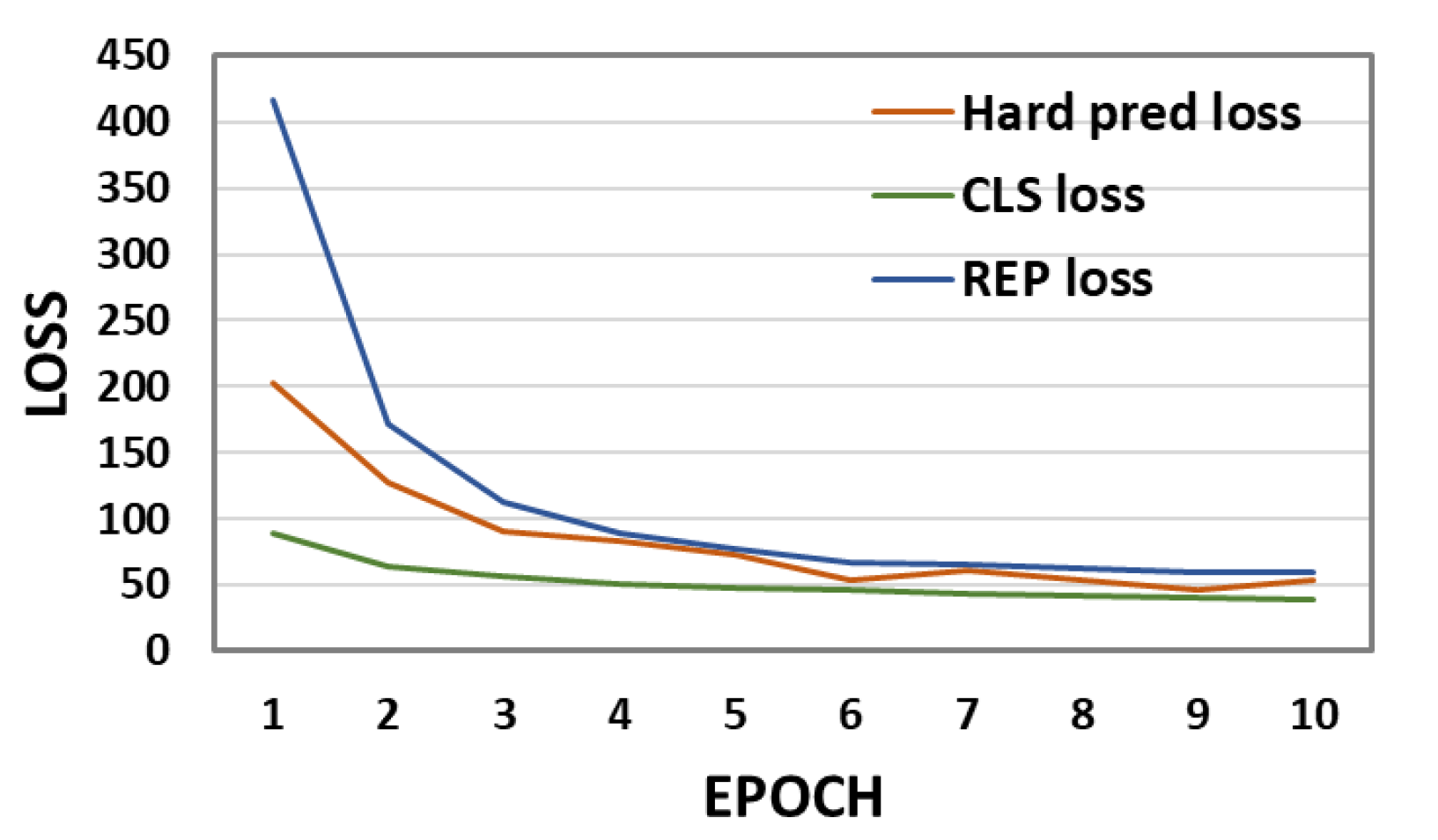}}
    \subfigure[TRMD-ColBERT]{\includegraphics[width=0.48\linewidth]{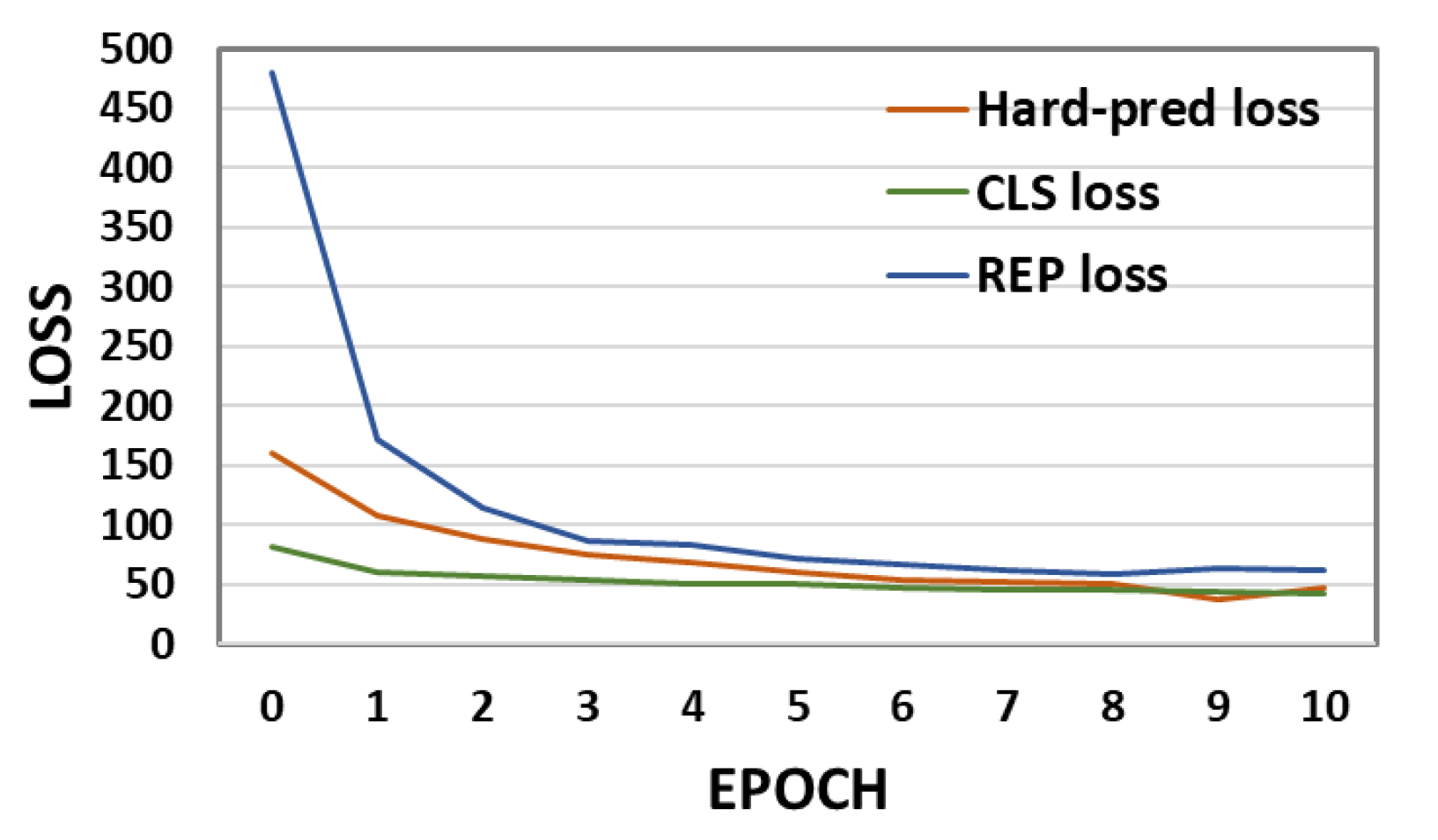}}
    \caption{Convergence of losses for Robust04 dataset.}
    \label{fig:distil_loss}
\end{figure}

\subsection{Result and Analysis}
Table \ref{tab:result0} shows the overall result of the experiment. We first compare the performances of cross-encoder monoBERT and bi-encoder TwinBERT and ColBERT. Comparing P@20, monoBERT shows better performance than the two bi-encoder models by more than 20\% on Robust04 and Clueweb09. This result confirms that bi-encoder NRMs generally perform worse than cross-encoder NRMs, supporting the need to improve the performance of bi-encoder models.

We then examine the change of each loss in TRMD to check whether the teachers effectively distill knowledge to the student. Figure \ref{fig:distil_loss} shows that three losses, $\mathcal{L}_{CLS}$, $\mathcal{L}_{REP}$, and $\mathcal{L}_{hard-pred}$, all decrease as the epoch increases. This result demonstrates that the learning dynamics of TRMD works well as we intended and that information in the representation of both teachers is effectively distilled to the student model through TRMD.

Next, we compare the performances of TRMD models with the baseline models. In Table \ref{tab:result0}, P@20 score of TRMD-TwinBERT on Robust04 is approximately 11.4\% higher than that of TwinBERT, with $p$-value smaller than 0.01. TRMD-ColBERT also outperforms ColBERT by more than 6.0\% on Robust04. Experiments on Clueweb09b show similar results with 4.6\% and 5.4\% improvements respectively, ensuring that TRMD significantly improves the ranking performance of the bi-encoder NRMs.


Finally, we compare the performance of TRMD and TR, where TR uses two rankers without any distillation, as an ablation study. The performance of P@20 on Robust04 shows that TR-TwinBERT is not significantly better than TwinBERT. The performance of TR-TwinBERT is even worse than that of TwinBERT when tested on the Clueweb09b dataset, indicating that simply using two rankers without knowledge distillation does not improve performance.

We suggest a possible explanation on why TRMD improves the performance of the bi-encoder models. Through TRMD, the student model is trained to learn from a bi-encoder teacher and a cross-encoder teacher. Figure \ref{fig:distil_loss} confirms that the student model successfully learns from both of the teachers. Between the two teachers, the cross-encoder teacher easily outperforms the bi-encoder teacher because the BERT part of a cross-encoder is allowed to utilize the relationship between query and document. Then, the cross-encoder's BERT representation can contain useful information on the relationship between query and document. Our results indicate that the student bi-encoder has learned to utilize the relationship by distilling the cross-encoder teacher's BERT representation. Simply having two rankers is not sufficient as we can tell from the performance of TR-TwinBERT and TR-ColBERT. Ensemble effect cannot explain the gain of TRMD, either. The cross-encoder teacher's output cannot be directly utilized as an input of an ensemble model because the student is constrained to be a bi-encoder. Therefore, our results indicate that even a bi-encoder can learn useful rules and patterns that are related to the relationship between query and document by having a powerful cross-encoder as a teacher. 


We mainly implement TRMD for improving bi-encoder models, but it is also possible to apply the TRMD approach to the models with cross-encoder structures.
We experimented TRMD-TwinBERT and TRMD-ColBERT after changing the bi-encoder students into a cross-encoder monoBERT student.
As shown at the bottom of Table \ref{tab:result0}, TRMD-TwinBERT using monoBERT as a student outperforms monoBERT by more than 4\% on Clueweb09b, and other cross-encoder TRMDs also outperform monoBERT.
These results demonstrate that TRMD can be used to enhance the performance of cross-encoder NRMs as well.

\section{Conclusion}
In this paper, we propose TRMD, a method enabling a bi-encoder model to learn from cross-encoder and bi-encoder teachers by applying multi-teacher knowledge distillation. We train a student model with three different losses(i.e., hard prediction, CLS, and representation losses) to enhance the performance. 
We verify that the improvement comes from the distillation process
by comparing TRMD with TR. 
Building a high-performing neural ranking model with less computation is an important research topic in the IR community. We believe that TRMD can effectively use knowledge distillation to improve bi-encoder neural ranking models.

\begin{acks}
This work was supported in part by Naver corporation (`Development of information retrieval system with large language models') and in part by the NRF grant (NRF-2020R1A2C2007139) funded by the Korea Government (MSIT).
\end{acks}

\bibliographystyle{ACM-Reference-Format}
\bibliography{acmart}


\begin{thebibliography}{14}


\ifx \showCODEN    \undefined \def \showCODEN     #1{\unskip}     \fi
\ifx \showDOI      \undefined \def \showDOI       #1{#1}\fi
\ifx \showISBNx    \undefined \def \showISBNx     #1{\unskip}     \fi
\ifx \showISBNxiii \undefined \def \showISBNxiii  #1{\unskip}     \fi
\ifx \showISSN     \undefined \def \showISSN      #1{\unskip}     \fi
\ifx \showLCCN     \undefined \def \showLCCN      #1{\unskip}     \fi
\ifx \shownote     \undefined \def \shownote      #1{#1}          \fi
\ifx \showarticletitle \undefined \def \showarticletitle #1{#1}   \fi
\ifx \showURL      \undefined \def \showURL       {\relax}        \fi
\providecommand\bibfield[2]{#2}
\providecommand\bibinfo[2]{#2}
\providecommand\natexlab[1]{#1}
\providecommand\showeprint[2][]{arXiv:#2}

\bibitem[\protect\citeauthoryear{Dehghani, Zamani, Severyn, Kamps, and
  Croft}{Dehghani et~al\mbox{.}}{2017}]%
        {dehghani2017neural}
\bibfield{author}{\bibinfo{person}{Mostafa Dehghani}, \bibinfo{person}{Hamed
  Zamani}, \bibinfo{person}{Aliaksei Severyn}, \bibinfo{person}{Jaap Kamps},
  {and} \bibinfo{person}{W~Bruce Croft}.} \bibinfo{year}{2017}\natexlab{}.
\newblock \showarticletitle{Neural ranking models with weak supervision}. In
  \bibinfo{booktitle}{\emph{Proceedings of the 40th International ACM SIGIR
  Conference on Research and Development in Information Retrieval}}.
  \bibinfo{pages}{65--74}.
\newblock


\bibitem[\protect\citeauthoryear{Fukuda, Suzuki, Kurata, Thomas, Cui, and
  Ramabhadran}{Fukuda et~al\mbox{.}}{2017}]%
        {fukuda2017efficient}
\bibfield{author}{\bibinfo{person}{Takashi Fukuda}, \bibinfo{person}{Masayuki
  Suzuki}, \bibinfo{person}{Gakuto Kurata}, \bibinfo{person}{Samuel Thomas},
  \bibinfo{person}{Jia Cui}, {and} \bibinfo{person}{Bhuvana Ramabhadran}.}
  \bibinfo{year}{2017}\natexlab{}.
\newblock \showarticletitle{Efficient Knowledge Distillation from an Ensemble
  of Teachers.}. In \bibinfo{booktitle}{\emph{Interspeech}}.
  \bibinfo{pages}{3697--3701}.
\newblock


\bibitem[\protect\citeauthoryear{He, Zhang, Ren, and Sun}{He
  et~al\mbox{.}}{2016}]%
        {he2016deep}
\bibfield{author}{\bibinfo{person}{Kaiming He}, \bibinfo{person}{Xiangyu
  Zhang}, \bibinfo{person}{Shaoqing Ren}, {and} \bibinfo{person}{Jian Sun}.}
  \bibinfo{year}{2016}\natexlab{}.
\newblock \showarticletitle{Deep residual learning for image recognition}. In
  \bibinfo{booktitle}{\emph{Proceedings of the IEEE conference on computer
  vision and pattern recognition}}. \bibinfo{pages}{770--778}.
\newblock


\bibitem[\protect\citeauthoryear{Hinton, Vinyals, and Dean}{Hinton
  et~al\mbox{.}}{2015}]%
        {hinton2015distilling}
\bibfield{author}{\bibinfo{person}{Geoffrey Hinton}, \bibinfo{person}{Oriol
  Vinyals}, {and} \bibinfo{person}{Jeff Dean}.}
  \bibinfo{year}{2015}\natexlab{}.
\newblock \showarticletitle{Distilling the Knowledge in a Neural Network}.
\newblock \bibinfo{journal}{\emph{stat}}  \bibinfo{volume}{1050}
  (\bibinfo{year}{2015}), \bibinfo{pages}{9}.
\newblock


\bibitem[\protect\citeauthoryear{Hui, Yates, Berberich, and De~Melo}{Hui
  et~al\mbox{.}}{2018}]%
        {hui2018co}
\bibfield{author}{\bibinfo{person}{Kai Hui}, \bibinfo{person}{Andrew Yates},
  \bibinfo{person}{Klaus Berberich}, {and} \bibinfo{person}{Gerard De~Melo}.}
  \bibinfo{year}{2018}\natexlab{}.
\newblock \showarticletitle{Co-PACRR: A context-aware neural IR model for
  ad-hoc retrieval}. In \bibinfo{booktitle}{\emph{Proceedings of the eleventh
  ACM international conference on web search and data mining}}.
  \bibinfo{pages}{279--287}.
\newblock


\bibitem[\protect\citeauthoryear{Humeau, Shuster, Lachaux, and Weston}{Humeau
  et~al\mbox{.}}{2020}]%
        {humeau2019poly}
\bibfield{author}{\bibinfo{person}{Samuel Humeau}, \bibinfo{person}{Kurt
  Shuster}, \bibinfo{person}{Marie-Anne Lachaux}, {and} \bibinfo{person}{Jason
  Weston}.} \bibinfo{year}{2020}\natexlab{}.
\newblock \showarticletitle{Poly-encoders: Architectures and Pre-training
  Strategies for Fast and Accurate Multi-sentence Scoring}. In
  \bibinfo{booktitle}{\emph{International Conference on Learning
  Representations}}.
\newblock


\bibitem[\protect\citeauthoryear{Huston and Croft}{Huston and Croft}{2014}]%
        {huston2014parameters}
\bibfield{author}{\bibinfo{person}{Samuel Huston} {and}
  \bibinfo{person}{W~Bruce Croft}.} \bibinfo{year}{2014}\natexlab{}.
\newblock \showarticletitle{Parameters learned in the comparison of retrieval
  models using term dependencies}.
\newblock \bibinfo{journal}{\emph{Ir, University of Massachusetts}}
  (\bibinfo{year}{2014}).
\newblock


\bibitem[\protect\citeauthoryear{Khattab and Zaharia}{Khattab and
  Zaharia}{2020}]%
        {khattab2020colbert}
\bibfield{author}{\bibinfo{person}{Omar Khattab} {and} \bibinfo{person}{Matei
  Zaharia}.} \bibinfo{year}{2020}\natexlab{}.
\newblock \showarticletitle{Colbert: Efficient and effective passage search via
  contextualized late interaction over bert}. In
  \bibinfo{booktitle}{\emph{Proceedings of the 43rd International ACM SIGIR
  Conference on Research and Development in Information Retrieval}}.
  \bibinfo{pages}{39--48}.
\newblock


\bibitem[\protect\citeauthoryear{Kingma and Ba}{Kingma and Ba}{2015}]%
        {kingma2015adam}
\bibfield{author}{\bibinfo{person}{Diederik~P Kingma} {and}
  \bibinfo{person}{Jimmy Ba}.} \bibinfo{year}{2015}\natexlab{}.
\newblock \showarticletitle{Adam: A Method for Stochastic Optimization}. In
  \bibinfo{booktitle}{\emph{ICLR (Poster)}}.
\newblock


\bibitem[\protect\citeauthoryear{Lu, Jiao, and Zhang}{Lu et~al\mbox{.}}{2020}]%
        {lu2020twinbert}
\bibfield{author}{\bibinfo{person}{Wenhao Lu}, \bibinfo{person}{Jian Jiao},
  {and} \bibinfo{person}{Ruofei Zhang}.} \bibinfo{year}{2020}\natexlab{}.
\newblock \showarticletitle{TwinBERT: Distilling Knowledge to Twin-Structured
  Compressed BERT Models for Large-Scale Retrieval}. In
  \bibinfo{booktitle}{\emph{Proceedings of the 29th ACM International
  Conference on Information \& Knowledge Management}}.
  \bibinfo{pages}{2645--2652}.
\newblock


\bibitem[\protect\citeauthoryear{MacAvaney, Yates, Cohan, and
  Goharian}{MacAvaney et~al\mbox{.}}{2019}]%
        {macavaney2019cedr}
\bibfield{author}{\bibinfo{person}{Sean MacAvaney}, \bibinfo{person}{Andrew
  Yates}, \bibinfo{person}{Arman Cohan}, {and} \bibinfo{person}{Nazli
  Goharian}.} \bibinfo{year}{2019}\natexlab{}.
\newblock \showarticletitle{CEDR: Contextualized embeddings for document
  ranking}. In \bibinfo{booktitle}{\emph{Proceedings of the 42nd International
  ACM SIGIR Conference on Research and Development in Information Retrieval}}.
  \bibinfo{pages}{1101--1104}.
\newblock


\bibitem[\protect\citeauthoryear{Mitra, Craswell, et~al\mbox{.}}{Mitra
  et~al\mbox{.}}{2018}]%
        {mitra2018introduction}
\bibfield{author}{\bibinfo{person}{Bhaskar Mitra}, \bibinfo{person}{Nick
  Craswell}, {et~al\mbox{.}}} \bibinfo{year}{2018}\natexlab{}.
\newblock \showarticletitle{An introduction to neural information retrieval}.
\newblock \bibinfo{journal}{\emph{Foundations and Trends{\textregistered} in
  Information Retrieval}} \bibinfo{volume}{13}, \bibinfo{number}{1}
  (\bibinfo{year}{2018}), \bibinfo{pages}{1--126}.
\newblock


\bibitem[\protect\citeauthoryear{Nogueira, Cho, and Scholar}{Nogueira
  et~al\mbox{.}}{2019}]%
        {nogueira2019passage}
\bibfield{author}{\bibinfo{person}{Rodrigo Nogueira},
  \bibinfo{person}{Kyunghyun Cho}, {and} \bibinfo{person}{CIFAR Azrieli~Global
  Scholar}.} \bibinfo{year}{2019}\natexlab{}.
\newblock \showarticletitle{PASSAGE RE-RANKING WITH BERT}.
\newblock \bibinfo{journal}{\emph{arXiv preprint arXiv:1901.04085}}
  (\bibinfo{year}{2019}).
\newblock


\bibitem[\protect\citeauthoryear{You, Xu, Xu, and Tao}{You
  et~al\mbox{.}}{2017}]%
        {you2017learning}
\bibfield{author}{\bibinfo{person}{Shan You}, \bibinfo{person}{Chang Xu},
  \bibinfo{person}{Chao Xu}, {and} \bibinfo{person}{Dacheng Tao}.}
  \bibinfo{year}{2017}\natexlab{}.
\newblock \showarticletitle{Learning from multiple teacher networks}. In
  \bibinfo{booktitle}{\emph{Proceedings of the 23rd ACM SIGKDD International
  Conference on Knowledge Discovery and Data Mining}}.
  \bibinfo{pages}{1285--1294}.
\newblock


\end{thebibliography}


\end{document}